\documentclass[a4paper,11pt]{article}
\usepackage{pos}

\usepackage{xspace}
\usepackage{hyperref}
\usepackage{wrapfig}
\usepackage{enumitem}
\usepackage{lineno}

\newcommand{\Xmax}{\ensuremath{X_\text{max}}\xspace}
\newcommand{\sib}[1]{\textsc{Sibyll}\,#1\xspace}

\newcommand{\qgsii}{\textsc{QGSJet~II-04}\xspace}

\newcommand{\eposlhc}{\textsc{Epos-LHC}\xspace}

\newcommand{\gcm}{\,\ensuremath{\text{g}/\text{cm}^{2}}}

\newcommand{\DeltaXmax}{\ensuremath{\Delta\Xmax}\xspace}

\title{A Data-driven Heavy-Metal Scenario for Ultra-High-Energy Cosmic Rays}

\author[a]{Jakub Vícha} 
\author*[a]{Alena Bakalová}
\author[a]{Ana L. Müller}
\author[a]{Olena Tkachenko}
\author[b, c]{Maximilian K. Stadelmaier}

\affiliation[a]{Institute of Physics of the Czech Academy of Sciences, Prague, Czech Republic}
\affiliation[b] {Universita degli Studi di Milano, Dipartimento di Fisica \& INFN, Sezione di Milano, Milano, Italy}
\affiliation[c] {Karlsruhe Institute of Technology, Institut fur Astroteilchenphysik, Karlsruhe, Germany}

\emailAdd{bakalova@fzu.cz}

\abstract{The mass composition of ultra-high-energy cosmic rays (UHECRs) is usually inferred from the depth of the shower maximum (\Xmax) of cosmic-ray showers, which is only ambiguously determined by modern hadronic interaction models. We present a data-driven interpretation of UHECRs, the \textit{heavy-metal scenario}, which assumes pure iron nuclei above $10^{19.6}$~eV ($\approx 40$~EeV) as the heaviest observed mass composition and introduces a global shift in the \Xmax scale predicted by the two hadronic interaction models \qgsii and \sib{2.3d}. We investigate the consequences of the proposed mass-composition model based on the obtained shifts in the \Xmax values, which naturally lead to a heavier mass composition of UHECRs than conventionally assumed.  We explore the consequences of our model on the energy evolution of relative fractions of primary species, consequently decomposed energy spectrum, hadronic-interaction studies and the arrival directions of UHECRs. We show that within this scenario, presented recently in \cite{Vicha_2025}, the cosmic-ray measurements can be interpreted in a more consistent way.

}

\ConferenceLogo{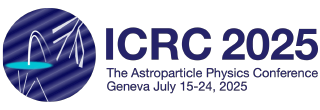}

\FullConference{39th International Cosmic Ray Conference (ICRC2025)\\
 15–24 July 2025\\
Geneva, Switzerland\\}

\begin{document}
\maketitle

\section{Introduction}

Understanding the mass composition of ultra-high-energy cosmic rays (UHECRs) is an important step towards determining their sources. In addition, UHECRs provide a unique opportunity to study hadronic interactions at center-of-mass energies (100s\,TeV) far beyond the reach of current human-made accelerators (13\,TeV). The mass composition of UHECRs is commonly estimated from the depth of the shower maximum (\Xmax) and the number of muons produced during the development of extensive air showers that is detected at the ground level.

However, comparisons of these observables with model predictions reveal a tension. Recent results of the Pierre Auger Observatory show a discrepancy between the measured data and the models at more than 5$\sigma$ \cite{PierreAuger:2024neu}. This study suggests that a consistent description of the measured data requires a shift in the \Xmax scale by $\approx 20~\rm{g/cm}^2 - 50~\rm{g/cm}^2$ together with rescaling of the hadronic component of the shower by $\approx 15\% - 25\%$ in the models \eposlhc \cite{EposLHC}, \qgsii \cite{Qgsjet} and \sib{2.3d} \cite{Sibyll}, which reduces the previous so-called muon puzzle approximately to its half. Moreover, machine-learning-based analyses of the Surface Detector data of the Pierre Auger Observatory \cite{AugerDNN:PRD} recently provided a precise estimation of the mean and fluctuation of \Xmax, for the first time, for energies from 3~EeV up to 100~EeV. Interpreting these data with \qgsii leads to nonphysical negative values of the variance of logarithmic atomic mass $\ln A$, $\sigma^2(\ln A)$. For the other two models, \sib{2.3d} and \eposlhc, $\sigma^2(\ln A)\approx 0$ suggests a pure composition, while the first moment of $\ln A$ suggests a continuous increase of $\ln A$ with energy. 


We present the \textit{heavy-metal scenario}, very recently proposed in \cite{Vicha_2025}. This scenario is built on two simple assumptions. First, we assume that the model predictions of \Xmax can be shifted by an energy- and mass-independent value, while keeping the other predictions of the hadronic interaction models unchanged. Secondly, we assume pure iron nuclei above $10^{19.6}$~eV, where the fluctuations of \Xmax are consistent with predictions for pure iron nuclei. We analyze publicly-available data from the Pierre Auger Observatory and investigate the implications of this scenario, which, from definition, leads to a heavier mass composition of UHECRs over a wide range of energies than conventionally interpreted. We explore how such an extreme mass-composition scenario affects the key aspects of UHECRs interpretations, including the inferred mass composition and spectral features, the muon puzzle, and arrival direction studies. For a more comprehensive analysis and discussions, we refer the reader to the full publication \cite{Vicha_2025}.

\section{Adjustment of the \Xmax scale}

\begin{figure*}
    \includegraphics[width=0.46\textwidth]{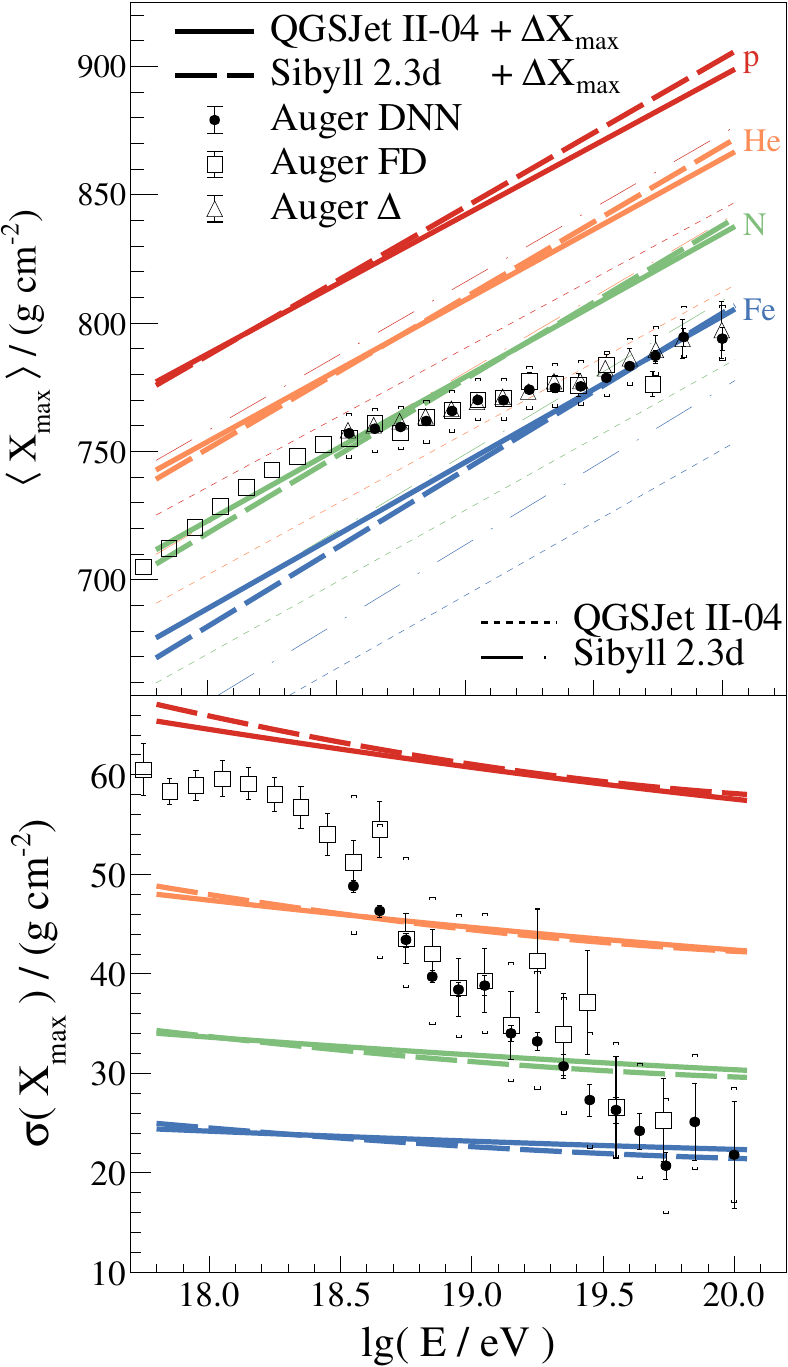}
    \hspace{0.2cm}
    \includegraphics[width=0.5\textwidth]{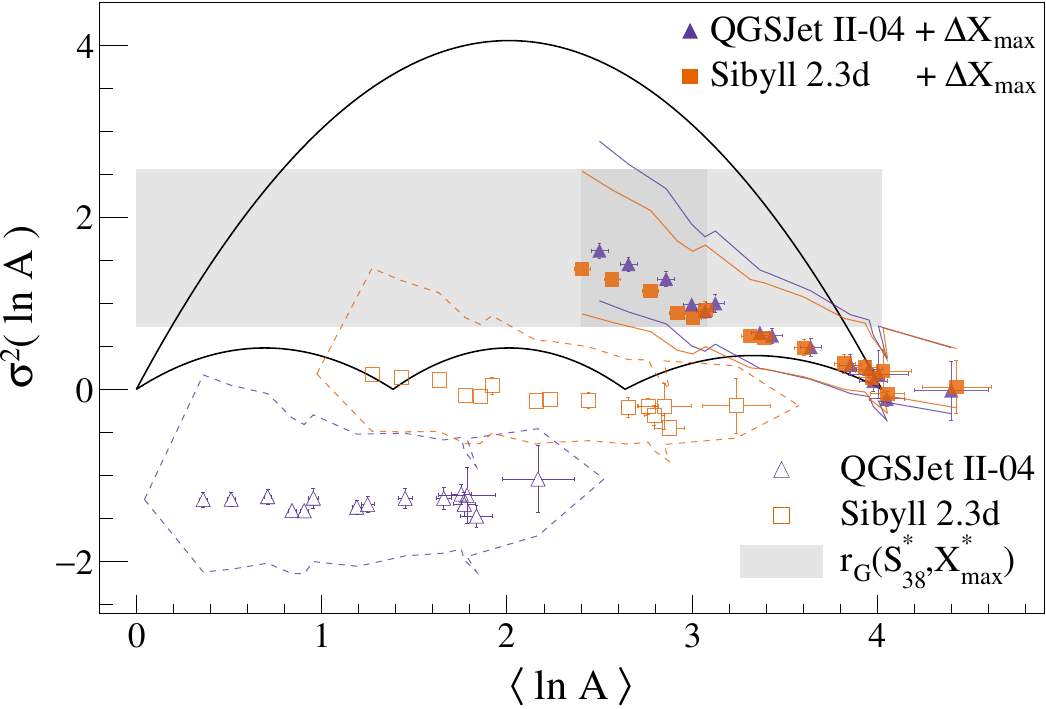}
    \caption{\textit{Left panel}: The energy evolution of the mean and standard deviation of $X_{\rm{max}}$ for data \citep{PierreAuger:2024flk,XmaxICRC19,DeltaICRC19} measured by the Pierre Auger Observatory (black). Thin lines show the original model predictions for four primary species, thick lines represent the adjusted model predictions. \textit{Right panel}: The interpreted mean and variance of $\ln A$ from the Auger DNN measurement for unmodified (open markers) and modified (full markers) model predictions. The region including the possible combinations of p, He, N and Fe nuclei is indicated by black curves. Taken from \cite{Vicha_2025}.}
    \label{fig:XmaxScaleFit}
 \end{figure*}

Assuming pure iron nuclei above $10^{19.6}$~eV ($\approx40$\,EeV) while keeping the elongation rate unchanged, the shift of the predicted \Xmax scale, \DeltaXmax, is fitted using the Auger DNN data \cite{AugerDNN:PRD}. The obtained shifts are $\DeltaXmax=52\pm1^{+11}_{-8}\,\gcm$ and $\DeltaXmax=29\pm1^{+12}_{-7}\,\gcm$ for \qgsii and \sib{2.3d}, respectively. Note that these values are consistent with the findings from \cite{PierreAuger:2024neu} at 3\,EeV-10\,EeV. The mean and standard deviation of the \Xmax from the Fluorescence Detector measurements \cite{XmaxICRC19} and from the Surface Detector measurements, the DNN method \cite{,AugerDNN:PRD} and the $\Delta$-method \cite{DeltaICRC19}, are shown in the left panel of Fig.~\ref{fig:XmaxScaleFit}. The thin lines represent the predictions of the unmodified models of hadronic interactions and the thick lines the predictions corrected for \DeltaXmax. Interpreting the \Xmax moments using the moments of $\ln A$, the umbrella plot for the unmodified and modified model predictions is depicted in the right panel of Figure~\ref{fig:XmaxScaleFit}. While the original model predictions lie mostly outside of the allowed region for all possibilities when mixing four different primary species (protons, and helium, nitrogen and iron nuclei), the shifted model predictions are within the allowed region of the umbrella plot.
Interestingly, also the value of $\sigma(\ln A)$ inferred from the correlation coefficient $r_\text{G}$ between the ground signal and \Xmax \cite{MixedAnkle} is consistent with the shifted models.

\section{Mass Composition and Energy Spectrum}

\begin{figure*}
    \includegraphics[width=0.46\textwidth]{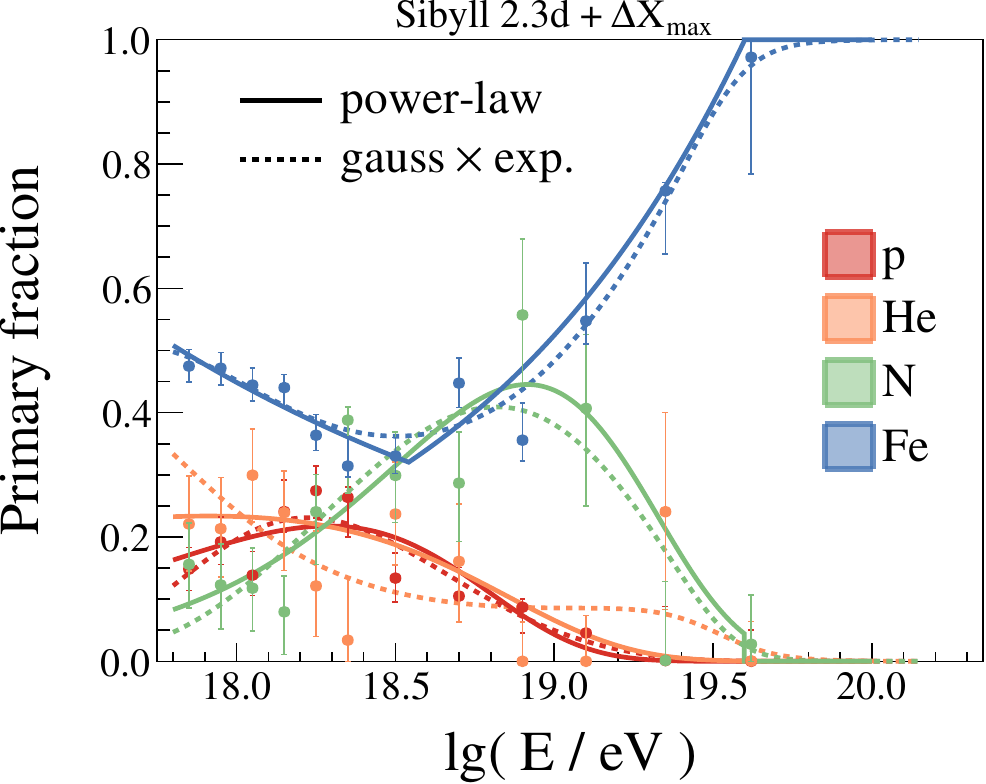}
    \hspace{0.2cm}
    \includegraphics[width=0.46\textwidth]{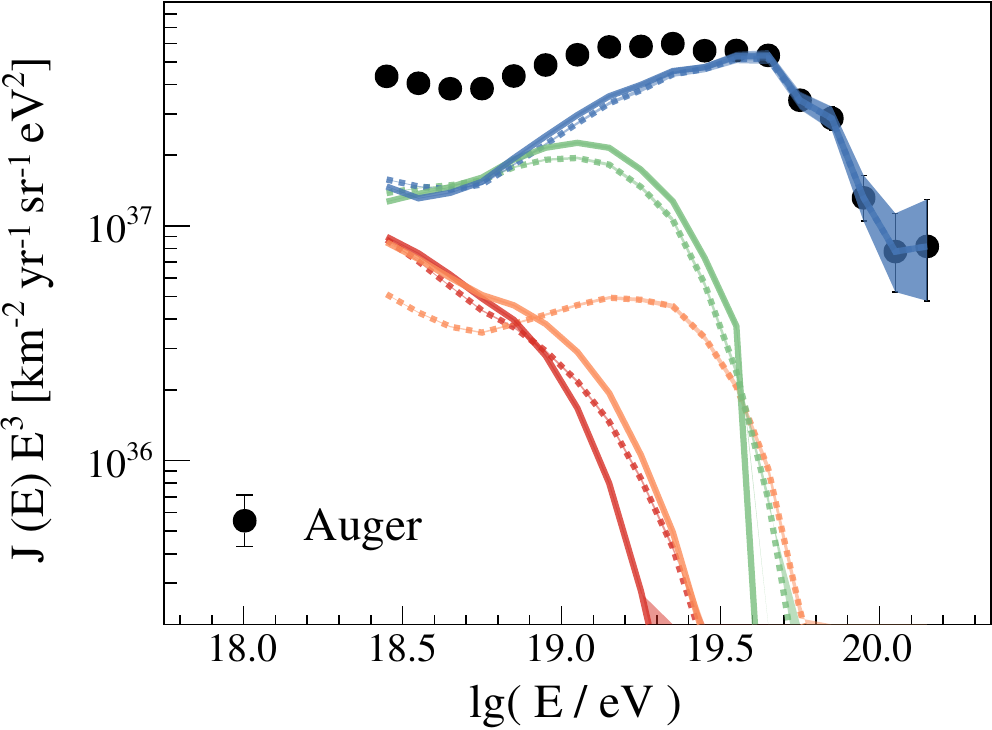}
    \caption{\textit{Left panel}: Energy evolutions of the primary fractions for \sib{2.3d}+\DeltaXmax from fits to the \Xmax distributions \cite{Auger-LongXmaxPaper}. The full and dotted lines represent two parameterizations of the fraction evolutions. \textit{Right panel}: Differential fluxes of individual primary species as a function of energy for \sib{2.3d}+\DeltaXmax obtained from the all-particle spectrum (black markers) from \cite{SDEnergySpectrum2020}. Taken from \cite{Vicha_2025}.}
    \label{fig:MassAndSpectrum}
 \end{figure*}

To obtain the energy evolution of the relative abundance of the four particle species with the adjusted model predictions of \Xmax for \qgsii and \sib{2.3d}, we fit the \Xmax distributions from \cite{Auger-LongXmaxPaper}. The fitted primary fractions for \sib{2.3d} are shown in the left panel of Figure~\ref{fig:MassAndSpectrum} together with two parameterizations of the energy evolution of the primary fractions; a smoothed Gaussian multiplied by
an exponential function and power-law functions with a simple exponential cutoff (see~\cite{Vicha_2025} for details). The energy evolution of the primary fractions shows a transition from lighter to heavier nuclei with energy, similarly to \cite{Auger-LongXmaxPaperMass} but with a smaller contribution of light elements in the mix. 

The energy spectrum of individual primaries is depicted in the right panel of Figure~\ref{fig:MassAndSpectrum} for \sib{2.3d} using the all-particle spectrum from \cite{SDEnergySpectrum2020}. The instep feature, around $\approx 15$~EeV, is then connected to the transition between nitrogen and iron nuclei within the heavy-metal scenario. The flux suppression is by assumption dominated by iron nuclei. The rigidity cutoff for both nitrogen and iron nuclei coincides at $\approx10^{18.2}$~V, suggesting a common origin of these primaries.

\section{Hadronic Interactions}

\begin{figure*}
    \includegraphics[width=0.52\textwidth]{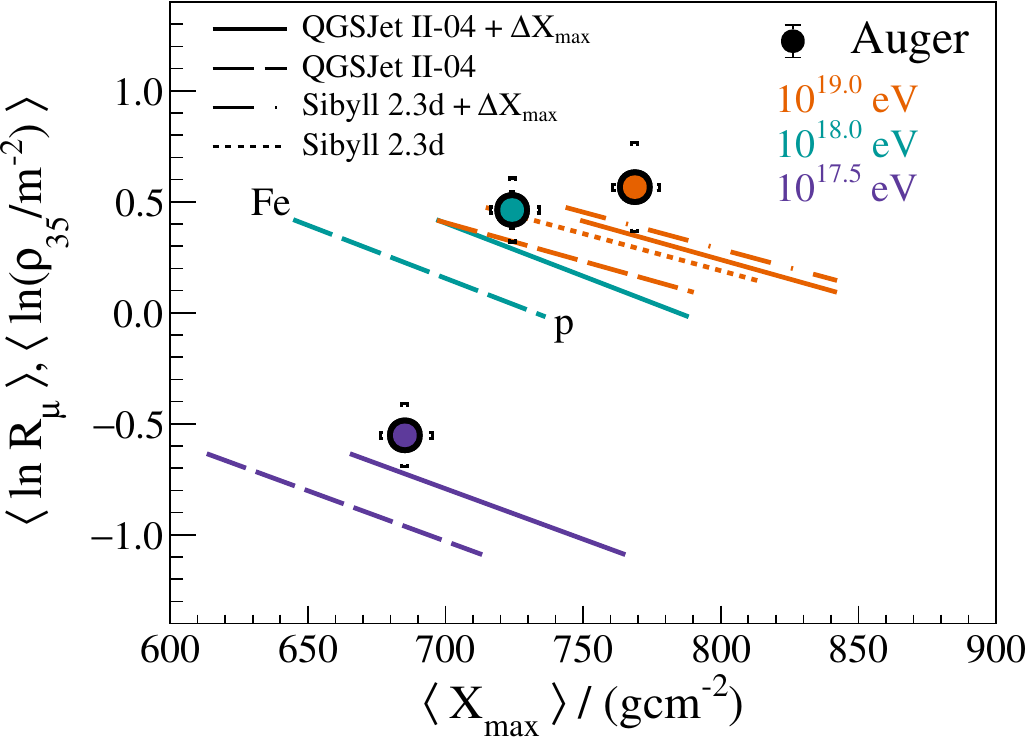}
    \hspace{0.2cm}
    \includegraphics[width=0.4\textwidth]{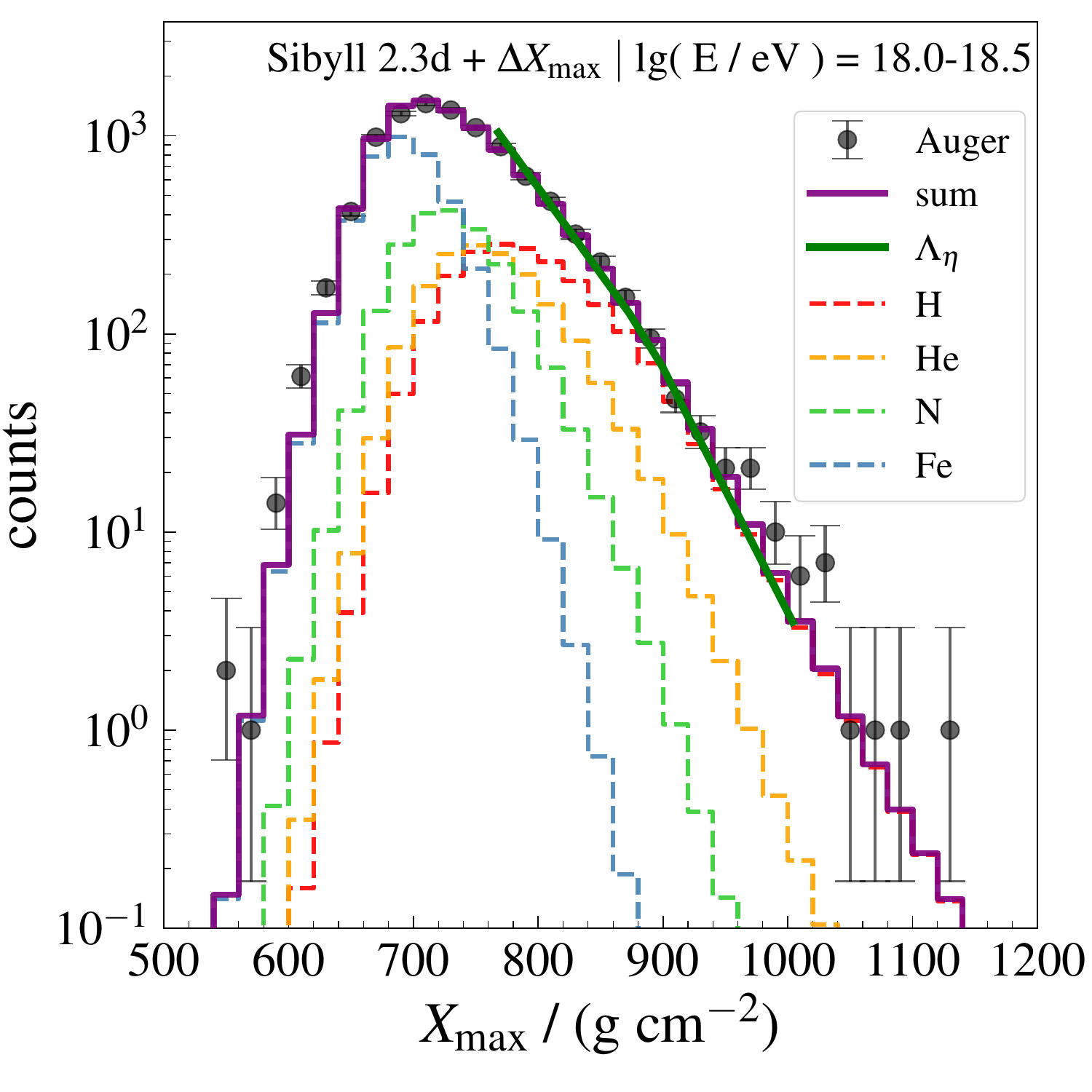}
    \caption{\textit{Left panel}: The muon number obtained by direct measurements at Auger~\citep{AmigaMuons,MuonFluct2020} compared to the predictions with and without the application of \DeltaXmax. \textit{Right panel}: The \Xmax distribution of Auger data from \citep{Auger-LongXmaxPaper} compared to the prediction for the heavy-metal scenario using \sib{2.3d}+\DeltaXmax in the energy bin $10^{18.0-18.5}$~eV. We illustrate the $\Lambda_\eta$ fit by the green line. Taken from \cite{Vicha_2025}.}
    \label{fig:Hadronics}
 \end{figure*}

The mass composition of cosmic rays is closely related to the number of muons in the extensive air showers. With the mass composition obtained within the heavy-metal scenario, we show the reduction of the so-called muon puzzle to approximately half in the left panel of Fig.~\ref{fig:Hadronics} in the direct muon measurements at the Pierre Auger Observatory when the \DeltaXmax is applied to the model predictions.
This result is consistent with the finding in \cite{PierreAuger:2024neu}.

The description of measured \Xmax distributions from \cite{Auger-LongXmaxPaper} is fair enough even for the \qgsii model, when \DeltaXmax is applied. We show in the right panel of Fig.~\ref{fig:Hadronics} such an example for $10^{18.0-18.5}$\,eV in the case of \sib{2.3d}+\DeltaXmax.
Estimating the slope of the \Xmax distribution as in \cite{ProtonAirXsec-Auger} ($\Lambda_\eta=(55.8\pm2.3\pm1.6)\gcm$), we obtain $\Lambda_\eta=(51.9\pm0.4)\gcm$ and $\Lambda_\eta=(50.0\pm0.4)\gcm$ in case of \qgsii+\DeltaXmax and \sib{2.3d}+\DeltaXmax, respectively. These somewhat lower values in case of the heavy-metal scenario might be a consequence of a higher helium contribution or a higher p-p cross-section or a lower elasticity that are extrapolated in the models from the accelerator measurements.

\section{Arrival directions}

We also investigate the implications of the heavy-metal scenario on the arrival directions of UHECRs, taking into account their deflections in the Galactic magnetic field (GMF). We use the eight UF23 models of the GMF \cite{UF23} together with the turbulent component of the Planck-tuned JF12 model \cite{JF12, JF12Planck}. 

\begin{figure*}
    \includegraphics[width=0.46\textwidth]{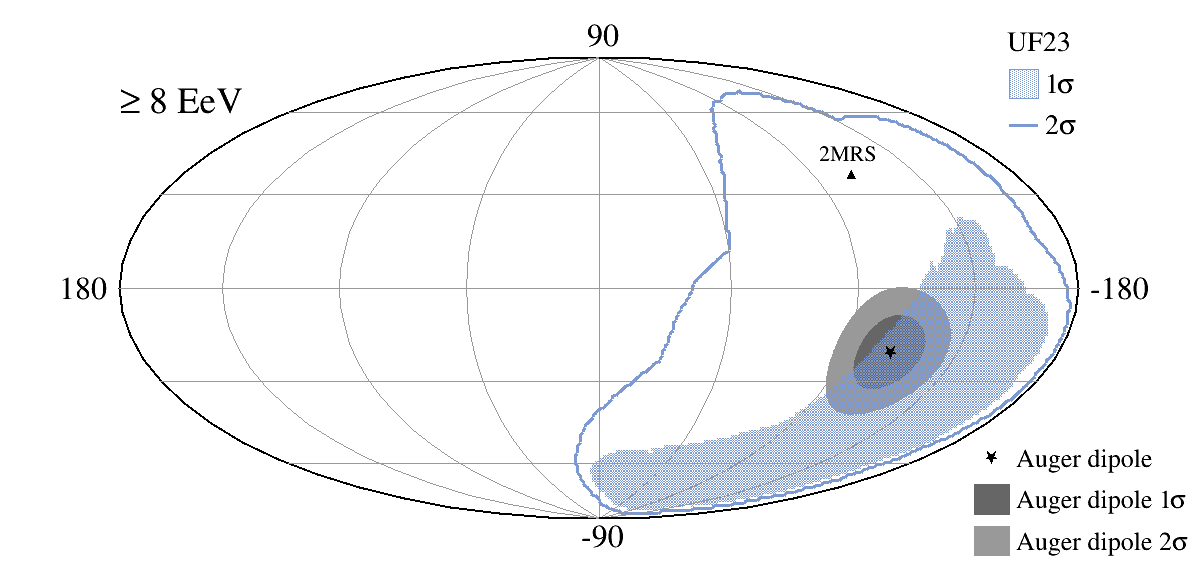}
    \hspace{0.2cm}
    \includegraphics[width=0.46\textwidth]{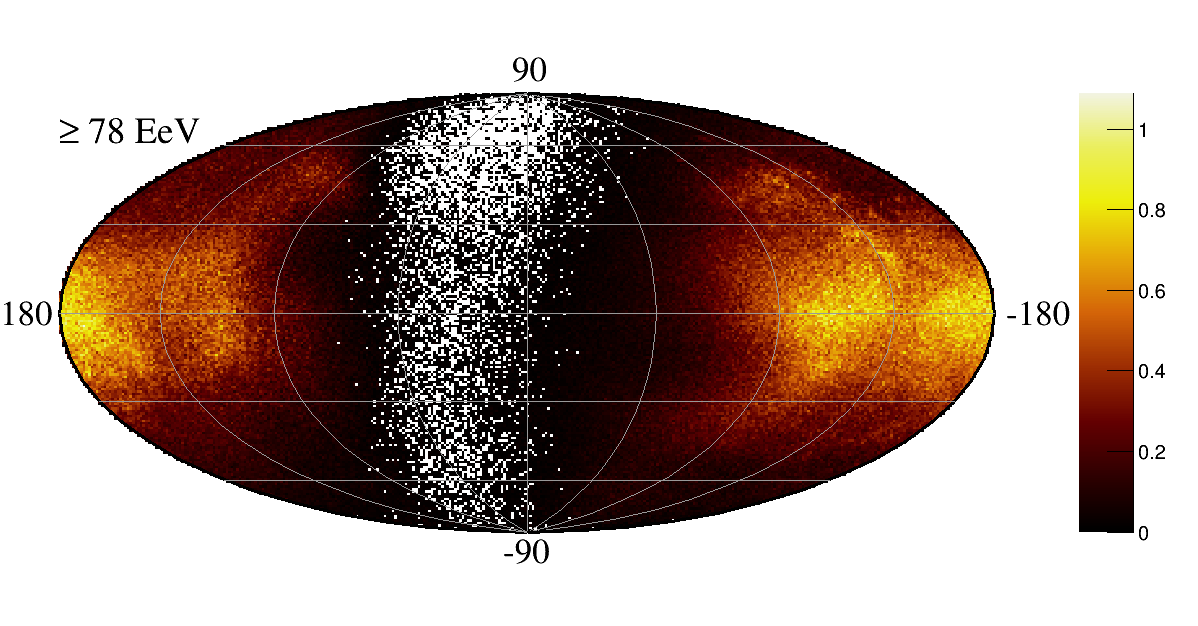}
    \caption{\textit{Left panel}: The $1\sigma$ and $2\sigma$ regions of allowed extragalactic dipole directions compatible with the measurement of the Pierre Auger Observatory \cite{AugerDipole24} after propagation in the GMF using UF23 models. \textit{Right panel}: The distribution of the backtracked directions at the edge of the Galaxy of the 100 most energetic Auger events above 78~EeV \cite{AugerMostEnergeticEvents} using UF23 models of the GMF. Taken from \cite{Vicha_2025}.}
    \label{fig:ADs}
 \end{figure*}

We first study the possible features of the extragalactic dipole that would result in the observed one on the Earth above 8~EeV \cite{AugerDipole24}. Following the same procedure as in \cite{BakalovaJCAP23}, we use the evolution of the mass composition of cosmic rays above 8 EeV obtained for \sib{2.3d}+\DeltaXmax. By simulating an ideal dipole flux outside the Galaxy with varying directions and amplitudes, we find regions of possible extragalactic dipole directions compatible with the Pierre Auger Observatory measurements \cite{AugerDipole24} at $1\sigma$ and $2\sigma$ levels. The $1\sigma$ and $2\sigma$ regions of the possible extragalactic directions of the dipole are visualized in the left panel of Figure~\ref{fig:ADs}. Due to the heavy mass composition, the extragalactic dipole amplitude must be relatively large ($\geq12\%$) to match the observed amplitude on the Earth. 

Secondly, we backtrack the 100 most energetic Auger events above 78~EeV \cite{AugerMostEnergeticEvents}, assuming pure iron nuclei. The simulated directions of the backtracked particles at the edge of the Galaxy are shown in the right panel of Figure~\ref{fig:ADs}. Due to the heavy mass composition, these particles experience strong deflections in the GMF, and the distribution of the backtracked cosmic rays at the edge of the Galaxy is the most dense in the direction close to the Galactic anti-center.

\section{Summary}

We present a data-driven mass composition scenario of ultra-high-energy cosmic rays, assuming a pure iron nuclei above the energy $10^{19.6}$~eV (approximately the region of steep flux suppression) and global shifts in the \Xmax scale of two hadronic interaction models \qgsii and \sib{2.3d} resulting in a consistent mass interpretation of the mean and variance of $\ln A$. The fitted \Xmax shifts are consistent with the results from \cite{PierreAuger:2024neu} obtained from the two-dimensional fits of \Xmax and the ground signal distributions at energies 3~EeV - 10~EeV, including the heavier mass composition than obtained for unmodified model predictions. Additionally, we show the consequences of the proposed heavy-metal scenario on the consistency of the decomposed energy spectrum, hadronic-interaction studies, and backtracked arrival directions in the Galactic magnetic field (GMF). 

Within the heavy-metal scenario, the decomposed energy fluxes of four primaries (protons and He, N and Fe nuclei) show the flux suppression of iron and nitrogen nuclei at rigidity $\approx 10^{18.2}$~V and explain the instep feature in the energy spectrum as a result of the fading of the nitrogen component. The agreement of the rigidity cutoff for nitrogen and iron nuclei also suggests their common origin. 
With the adjusted \Xmax scale and, therefore, heavier mass composition over the whole energy range, the muon puzzle is reduced by approximately half. The measured \Xmax distributions can be well described using the new mass-composition model. However, we obtain slightly lower values of the slope of the \Xmax distribution, that might be caused by a higher fraction of helium nuclei in our scenario or the need for a lower extrapolated p-p cross section or a higher elasticity in the modeled hadronic interactions. The observed dipole anisotropy above 8\,EeV is compatible with an extragalactic origin under our mass-composition scenario, requiring extragalactic dipole amplitudes $\geq12\%$. Assuming only iron nuclei, the arrival directions of the most energetic Auger events above 78\,EeV, when backtracked through the GMF, point towards the Galactic anti-center region, which is consistent with the expectations from isotropic arrival directions at the Earth. The estimated source luminosity points to only the most powerful astrophysical objects, such as hard X-ray AGNs, as viable candidates.

In the next steps, we plan to explore the possibility of including even heavier nuclei than iron at the highest energies, as suggested by recent studies, see e.g. \cite{FarrarBNS, ZhangUHUHECR}. Moreover, we intend to test the proposed scenario using next-generation hadronic interaction models that have been recently developed and made publicly available.

\section*{Acknowledgments}
The work was supported by the Czech Academy of Sciences: LQ100102401, Czech Science Foundation: 21-02226M, Ministry of Education, Youth and Sports, Czech Republic: FORTE CZ.02.01.01/00/22\_008/0004632, German Academic Exchange service (DAAD PRIME). 
The authors are very grateful to the Pierre Auger Collaboration for discussions about this work.

\bibliographystyle{JHEP}
\bibliography{biblio.bib}

\end{document}